\documentclass[twocolumns,9pt]{IEEEtran}
\usepackage{amsmath}
\usepackage{amsfonts}
\usepackage{mathrsfs}
\usepackage{pifont}
\usepackage{amssymb}
\usepackage{verbatim}
\usepackage{upgreek}
\usepackage{color}
\usepackage{graphicx}
\usepackage{algorithmic}
\usepackage{algorithm}
\usepackage{epsfig}

\usepackage{url}
\usepackage{multicol}
\usepackage{lipsum}
\usepackage{mathtools}
\usepackage{cuted}

\newcommand{\bzero}{\mbox{\boldmath{$0$}}}

\newcommand{\bA}{\mbox{\boldmath{$A$}}}

\newcommand{\be}{\mbox{\boldmath{$e$}}}

\newcommand{\bh}{\mbox{\boldmath{$h$}}}
\newcommand{\bI}{\mbox{\boldmath{$I$}}}

\newcommand{\bu}{\mbox{\boldmath{$u$}}}

\newcommand{\bv}{\mbox{\boldmath{$v$}}}
\newcommand{\bW}{\mbox{\boldmath{$W$}}}
\newcommand{\bw}{\mbox{\boldmath{$w$}}}

\newcommand{\bx}{\mbox{\boldmath{$x$}}}
\newcommand{\bY}{\mbox{\boldmath{$Y$}}}
\newcommand{\by}{\mbox{\boldmath{$y$}}}

\newcommand{\tr}{\mbox{\rm tr}\, }

\newtheorem{theorem}{Theorem}[section]

\newtheorem{proposition}[theorem]{Proposition}

\begin{document}

\title{Robust Downlink Transmit Optimization under Quantized Channel Feedback via the Strong Duality for QCQP}
\vspace{2cm}

\author{
Xianming Lin,~{\it Student Member, IEEE}, \thanks{This work was supported in part by the National Natural Science Foundation of China under Grants 11871168 and U1701266.}
Yongwei Huang,~{\it Senior Member, IEEE}, \thanks{X. Lin and Y. Huang are with School of Information Engineering, Guangdong University of Technology, University Town, Guangzhou, Guangdong 510006, China (email: ywhuang@gdut.edu.cn; 2111903006@mail2.gdut.edu.cn).}
Wing-Kin Ma,~{\it Fellow, IEEE}
\thanks{W.-K. Ma is with the Department of Electronic Engineering,
The Chinese University of Hong Kong, Shatin, Hong
Kong. (email: wkma@ieee.org).}
}

\maketitle

\begin{abstract}
Consider a robust multiple-input single-output downlink beamforming optimization problem in a frequency division duplexing system. The base station (BS) sends training signals to the users, and every user estimates the channel coefficients, quantizes the gain and the direction of the estimated channel and sends them back to the BS. Suppose that the channel state information at the transmitter is imperfectly known mainly due to the channel direction quantization errors, channel estimation errors and outdated channel effects. The actual channel is modeled as in an uncertainty set composed of two inequality homogeneous and one equality inhomogeneous quadratic constraints, in order to account for the aforementioned errors and effects. Then the transmit power minimization problem is formulated subject to robust signal-to-noise-plus-interference ratio constraints. Each robust constraint is transformed equivalently into a quadratic matrix inequality (QMI) constraint with respect to the beamforming vectors. The transformation is accomplished by an equivalent phase rotation process and the strong duality result for a quadratically constrained quadratic program. The minimization problem is accordingly turned into a QMI problem, and the problem is solved by a restricted linear matrix inequality relaxation with additional valid convex constraints. Simulation results are presented to demonstrate the performance of the proposed method, and show the efficiency of the restricted relaxation.
\end{abstract}



\section{Introduction}

In a multiuser multiple-input single-out (MISO) downlink communication system, linear beamforming technique has played a vital role in improving spectrum efficiency and alleviating mutual interference (see e.g. \cite{GSSBO09,Precoding-1,Precoding-2} and references therein). Among many existing optimal beamforming designs,  a quality-of-service (QoS) design problem is typically
formulated such that the power consumed in the base station (BS) is minimized subject to signal-to-interference-plus-noise ratio (SINR) constraints at the receivers. In this design, the BS must be able to obtain the channel state information at the transmitter (CSIT) and determine the SINRs. However, the BS often has only estimated and imperfect CSIT and thus the SINRs are computed in an inaccurate way. To address the inexactness, the presence of uncertainties in these estimates has to be taken into account. A prominent approach is to design the beamforming vectors such that they are robust against the CSIT imperfectness (see e.g. \cite{Ma2017-tsp} and references therein).

Therefore, robust beamforming optimization (based on the worst-case scenario) in different systems of communications has been widely studied in the past two decades (see e.g. \cite{shenouda-davidson07,Vucic2009-tsp,Shen2012-tsp,HuangPalomarZhang2013-tsp,RobustBF-1,RobustBF-2,RobustBF-3,RobustBF-4,RobustBF-5,RobustBF-6}). In the references, an uncertainty set of the actual channel is modeled as a ball or an ellipsoid around the channel estimate at the transmitter, and common robust optimization tools, like traditional $S$-lemmas \cite{S-lemma-survey}, are employed to solve the robust beamforming design problems.
Recently, an extended $S$-lemma is established \cite{MHMD2016tsp} to solve a robust beamforming problem in the context of a frequency division duplexing (FDD) system \cite{Grassmannian2003}. In that system, each receiver estimates its channel coefficients and feeds back a quantized version of the channel gain and direction to the BS. Assuming perfect channel state information at the receiver (CSIR), high resolution of the channel gain quantizer and no outdated channel effects between the BS and user $k$, the perturbation set of the actual channel required at the BS is modeled as one inhomogeneous equality  and one homogeneous inequality quadratic constraints for the channel direction quantization errors. With the channel perturbation set, robust SINR constraints can be defined. Applying the extended $S$-lemma, each robust SINR constraint is turned into a quadratic matrix inequality (QMI) constraint with respect to the beamforming vectors, and a transmit power minimization problem subject to robust SINR constraints is reformulated into a QMI problem, and finally the robust optimal beamforming problem is solved via linear matrix inequality (LMI) relaxation \cite{MHMD2016tsp}. 

In this paper, we assume that there exist channel estimation errors and outdated channel effects (which looks more practical and reasonable), in addition to the channel direction quantization errors. Thereby, the perturbation set of the actual channel is represented by two homogeneous inequalities and one inhomogeneous equality quadratic constraints to account for the aforementioned errors and effects, under the assumption of perfect CSIR and high resolution of the channel gain quantizer. Hence, the extended $S$-lemma is not applicable to the robust SINR constraints which are defined according to the new channel model. Resorting to a certain equivalent phase rotation process and the strong duality result for a quadratically constrained quadratic program (QCQP) with a small number of constraints \cite{bookchapter-2010}, 
the robust beamforming problem of the power minimization subject to the robust SINR constraints is transformed equivalently into a QMI problem. Then the QMI problem is solved by a restricted LMI relaxation with additional  valid convex constraints. Our simulation examples demonstrate the performance of the robust design, and show the efficiency of the restricted LMI relaxation. Note that herein the robust optimization tool applied is the strong duality of QCQP in the context of an FDD downlink system while in the aforementioned works the optimization tools are the traditional $S$-lemmas in the settings of other variant systems of communications.


\section{Signal Model and Problem Formulation}


Let us consider a $K$-user unicast MISO downlink, where a BS equipped with $N_t$ antennas sends independent messages to $K$ single antenna users.  The BS employs linear beamforming to construct the transmitted signal, $\bx=\sum_{k=1}^K\bw_ks_k$, where $s_k$ is the normalized symbol intended for user $k$, and $\bw_k\in\mathbb{C}^{N_t}$ is the associated beamforming vector. The signal received by the $k$th user is
\begin{equation}\label{receive-signal-at-user-k}
y_k=\bh_k^H\bw_ks_k+\sum_{j=1,j\ne k}^K\bh_k^H\bw_js_j+n_k,
\end{equation}
where $\bh_k\in\mathbb{C}^{N_t}$ stands for the channel between the BS and receiver $k$, and $n_k\in\mathbb{C}$ represents the additive zero mean circular complex Gaussian noise. Therefore, the SINR at user $k$ is given by
\begin{equation}\label{SINR-user-k}
\mbox{SINR}_k=\frac{|\bh_k^H\bw_k|^2}{\sum_{j\ne k}|\bh_k^H\bw_j|^2+\sigma_k^2},
\end{equation}
where $\sigma_k^2$ is the noise variance.

A basic optimal beamforming problem is formulated as a total transmit energy minimization problem subject to each SINR no less than a prefix threshold (see e.g. \cite{GSSBO09} and references therein). 
When only imperfect CSIT is available to the BS, a robust optimal transmit beamforming problem is taken into account. We select the worst-case channel model as our initial development. That is, the following problem formulation is considered:
\begin{equation}\label{Power-min-SINR-qcqp-robust-0}
\begin{array}[c]{cl}
\underset{\{\bw_k\}}{\sf{minimize}}  & \begin{array}[c]{c}\sum_{k=1}^K\bw_k^H\bw_k\end{array}\\
\sf{subject\;to}               &
\begin{array}[t]{l}
\mbox{SINR}_k\ge\gamma_k,\,\forall \bh_k\in{\cal R}_k,
 \end{array}
\end{array}
\end{equation}
where ${\cal R}_k$ contains all the possible values that $\bh_k$ can take on. 
Obviously, problem \eqref{Power-min-SINR-qcqp-robust-0} can be recast into:
\begin{equation}\label{Power-min-SINR-qcqp-robust}
\begin{array}[c]{cl}
\underset{\{\bw_k\}}{\sf{minimize}}  & \begin{array}[c]{c}\sum_{k=1}^K\bw_k^H\bw_k\end{array}\\
\sf{s.t.}               &
\begin{array}[t]{l}
\bh_k^H\tilde\bW_k\bh_k\ge\sigma_k^2,\,\forall \bh_k\in{\cal R}_k,
 \end{array}
\end{array}
\end{equation}
where
\begin{equation}\label{define-tilde-W}
\tilde\bW_k=\frac{1}{\gamma_k}\bw_k\bw_k^H-\sum_{j=1,j\ne k}^K\bw_j\bw_j^H,\,\forall k,
\end{equation}
(see e.g. \cite[problem (4)]{Ma2017-tsp}). Here $\gamma_k$ can be expressed as $2^{r_k}-1$, where $r_k>0$ is a preset achievable rate value of user $k$.

Suppose that the downlink is an FDD system with quasi-static channels. With structured vector quantization \cite{Grassmannian2003}, receiver $k$ estimates the channel based on training signals sent by the BS, separately quantizes the gain and ``direction" of the channel and sends them back to the BS. Precisely, if $\tilde{\bh}_k$ denotes the receiver's estimate, the receiver quantizes $\sqrt{\alpha_k}=\|\tilde{\bh}_k\|$ using a scalar quantizer, and quantizes $\tilde{\bh}_{n_k}=\tilde{\bh}_k/\|\tilde{\bh}_k\|$ using memoryless vector quantization over a Grassmannian  codebook \cite{Grassmannian2003}; in other words, if ${\cal C}_k=\{\bv_{k1},\bv_{k2},\cdots,\bv_{kM_k}\}$ represents a Grassmannian codebook of $M_k$ unit-norm vector elements in $\mathbb{C}^{N_t}$ for user $k$, then the codebook element that characterizes the direction of the channel is given by $\bh_{q_k} =\arg\max_{\bv\in{\cal C}_k} |\tilde{\bh}_{n_k}^H\bv|^2$. 

We assume that $\tilde{\bh}_k$ is estimated accurately (perfect CSIR), and  $\alpha_k$  is quantized at a high resolution. The transmitter's estimate of the channel, $\sqrt{\alpha_k}\bh_{q_k}$, is related to the actual channel by
\begin{equation}\label{the-actual-channel-vector}
\bh_k=\sqrt{\alpha_k}(\bh_{q_k}+\be_k)+\bu_k,
\end{equation}
where $\be_k$ is  the channel direction quantization error, and $\bu_k$ is contributed by channel estimation errors and outdated channel effects.

Given the nature of $\be_k$ and $\bu_k$,
the channel perturbation set is defined as:
\begin{eqnarray}\nonumber
{\cal E}_k^\prime &=&\{(\be_k,\bu_k)\in\mathbb{C}^{N_t}\times\mathbb{C}^{N_t}:\|\be_k\|\le\epsilon_k,\\ \label{uncertainty-set-channel}
& &\|\bh_{q_k}+\be_k\|=1,\,\|\bu_k\|\le\beta_k\}.
\end{eqnarray}
Particularly, the case of $\bu_k$ vanishing in \eqref{the-actual-channel-vector} is studied in \cite{MHMD2016tsp}, and therein the channel error set reduces to
\begin{equation}\label{uncertainty-set-channel-2016-tsp}
{\cal E}_k^{\prime\prime}=\{\be_k\in\mathbb{C}^{N_t}~:~\|\be_k\|\le\epsilon_k,\,\|\bh_{q_k}+\be_k\|=1\},
\end{equation}
i.e., setting $\beta_k=0$ in \eqref{uncertainty-set-channel}. From the difference between \eqref{uncertainty-set-channel} and \eqref{uncertainty-set-channel-2016-tsp}, it is observed that the extended $S$-lemma in \cite{MHMD2016tsp} cannot be applied herein any more, and we have to find another optimization tool to handle the new channel model.

From \eqref{the-actual-channel-vector} and \eqref{uncertainty-set-channel}, we can set
$\tilde\be_k=\bh_{q_k}+\be_k$,
and rewrite perturbation set \eqref{uncertainty-set-channel} in terms of $(\bh_k,\tilde\be_k)$ into:
\begin{eqnarray}\nonumber
{\cal E}_k&=&\{(\bh_k,\tilde\be_k)\in\mathbb{C}^{N_t}\times\mathbb{C}^{N_t}:\|\tilde\be_k-\bh_{q_k}\|\le\epsilon_k,\\ \label{uncertainty-set-channel-recast}
& &\|\tilde\be_k\|=1,\|\bh_k-\sqrt{\alpha_k}\tilde\be_k\|\le\beta_k\}.
\end{eqnarray}

In order to avoid trivial discussion, we assume, throughout the paper, that $\epsilon_k\ne0$, $\beta_k\ne0$, and $\alpha_k\ne0$.
Observe that $(\sqrt{\alpha_k}\bh_{q_k},\bh_{q_k})$ is an interior point of the uncertainty set ${\cal E}_k$ (due to $\|\bh_{q_k}\|=1$), and that it is sufficient to assume that $\epsilon_k\le 2$ since when $\epsilon_k>2$,  condition  $\|\tilde\be_k-\bh_{q_k}\|\le\epsilon_k$ is the same as $\|\tilde\be_k-\bh_{q_k}\|\le2$ (note that $\|\tilde\be_k\|=\|\bh_{q_k}\|=1$).

With ${\cal E}_k$ in hand, original robust transmit beamforming problem \eqref{Power-min-SINR-qcqp-robust} can be reexpressed into:
\begin{equation}\label{R-OBP-O-h-new-1-reformulation-1}
\begin{array}[c]{cl}
\underset{\displaystyle\{\bw_k\}}{\sf{minimize}}  & \begin{array}[c]{c}\sum_{k=1}^K\bw_k^H\bw_k \end{array}\\
\sf{subject\;to}                              &
\begin{array}[c]{lll}
\underset{\displaystyle(\bh_k,\tilde\be_k)\in{\cal E}_k}{\sf{minimize}}  & \bh_k^H\tilde\bW_k\bh_k  &\ge\sigma_k^2,\,\forall k.
\end{array}
\end{array}
\end{equation}


\section{Equivalent QMI Reformulation via the Strong Duality between a QCQP Problem and its Dual}


The QCQP problem  at the  left-hand side of the  constraint is rewritten as the minimization problem:
\begin{equation}\label{the-worst-SINR-rewrite-qcqp}
\begin{array}[c]{cl}
\underset{(\bh_k,\tilde\be_k)}{\sf{minimize}}  & \begin{array}[c]{c}\bh_k^H\tilde\bW_k\bh_k\end{array}\\
\sf{subject\;to}               &
\begin{array}[t]{l}
\|\tilde\be_k-\bh_{q_k}\|\le\epsilon_k\\
\|\tilde\be_k\|=1\\
\|\bh_k-\sqrt{\alpha_k}\tilde\be_k\|\le\beta_k.
 \end{array}
\end{array}
\end{equation}
It is a nonconvex inhomogeneous QCQP with three constraints. It follows from \cite{bookchapter-2010} that there is a positive duality gap duality between \eqref{the-worst-SINR-rewrite-qcqp} and its dual in general.  However, it is observed that the Hessian matrices in the first and the second constraints are the identity matrix, which allows us to close the duality gap by some manipulations.


Specifically, the first constraint can be reformulated into
\begin{equation}\label{1-constraint-reform}
\Re(\bh_{q_k}^H\tilde\be_k)\ge1-\frac{\epsilon_k^2}{2}.
\end{equation}
In other words, QCQP problem \eqref{the-worst-SINR-rewrite-qcqp} can be recast into
\begin{equation}\label{the-worst-SINR-rewrite-qcqp-1}
\begin{array}[c]{cl}
\underset{(\bh_k,\tilde\be_k)}{\sf{minimize}}  & \begin{array}[c]{c}\bh_k^H\tilde\bW\bh_k\end{array}\\
\sf{subject\;to}               &
\begin{array}[t]{l}
\Re(\bh_{q_k}^H\tilde\be_k)\ge1-\frac{\epsilon_k^2}{2}\\
\|\tilde\be_k\|=1\\
\|\bh_k-\sqrt{\alpha_k}\tilde\be_k\|\le\beta_k.
 \end{array}
\end{array}
\end{equation}
Then we claim the following proposition.

\begin{proposition}\label{two-qcqps-equival}
Suppose that $\epsilon_k\le\sqrt{2}$. Then QCQP problem \eqref{the-worst-SINR-rewrite-qcqp-1}  is equivalent to the following QCQP problem
\begin{equation}\label{the-worst-SINR-rewrite-qcqp-2}
\begin{array}[c]{cl}
\underset{(\bh_k,\tilde\be_k)}{\sf{minimize}}  & \begin{array}[c]{c}\bh_k^H\tilde\bW\bh_k\end{array}\\
\sf{subject\;to}               &
\begin{array}[t]{l}
|\bh_{q_k}^H\tilde\be_k|^2\ge(1-\frac{\epsilon_k^2}{2})^2\\
\|\tilde\be_k\|^2=1\\
\|\bh_k-\sqrt{\alpha_k}\tilde\be_k\|^2\le\beta_k^2,
 \end{array}
\end{array}
\end{equation}
in the sense that they share the same optimal value.
\end{proposition}

The proof mainly includes an equivalent phase rotation technique, which we thus omit.
Therefore, we have
\begin{equation}\label{three-problems-equal-values}
v^\star(\eqref{the-worst-SINR-rewrite-qcqp})=v^\star(\eqref{the-worst-SINR-rewrite-qcqp-1})=v^\star(\eqref{the-worst-SINR-rewrite-qcqp-2}),
\end{equation}
where $v^\star((\cdot))$ stands for the optimal value of $(\cdot)$.
Note that problem \eqref{the-worst-SINR-rewrite-qcqp-2} is homogeneous and it can be reformulated into:
\begin{equation}\label{the-worst-SINR-rewrite-qcqp-homog-matrix}
\begin{array}[c]{ll}
\underset{\by_k}{\sf{minimize}}  & \begin{array}[c]{c}\by_k^H\bA_{k0}\by_k \end{array}\\
\sf{subject\;to}               &
\begin{array}[t]{l}
\by_k^H\bA_{k1}\by_k\ge(1-\epsilon_k^2/2)^2\\
\by_k^H\bA_{k2}\by_k=1\\
\by_k^H\bA_{k3}\by_k\le\beta_k^2,\\
\end{array}
\end{array}
\end{equation}
where $\by_k=[\bh_k^H, \tilde\be_k^H]^H$, and
\begin{equation}\label{define-another-Ak0-Ak1}
\bA_{k0}=\left[\begin{array}{cc}\tilde\bW_k&\bzero\\ \bzero&\bzero \end{array}\right],\,\bA_{k1}=\left[\begin{array}{cc}\bzero&\bzero \\ \bzero&\bh_{q_k}\bh_{q_k}^H\end{array}\right],
\end{equation}
\begin{equation}\label{define-another-Ak2-Ak3}
\bA_{k2}=\left[\begin{array}{cc}\bzero&\bzero \\ \bzero&\bI\end{array}\right],\,\bA_{k3}=\left[\begin{array}{cc}\bI& -\sqrt{\alpha_k}\bI \\ -\sqrt{\alpha_k}\bI & \alpha_k\bI  \end{array}\right].
\end{equation}
Then, the SDP relaxation problem is
\begin{equation}\label{the-worst-SINR-rewrite-qcqp-homog-matrix-SDR}
\begin{array}[c]{ll}
\underset{\bY_k}{\sf{minimize}}  & \begin{array}[c]{c}\tr(\bA_{k0}\bY_k) \end{array}\\
\sf{subject\;to}               &
\begin{array}[t]{l}
\tr(\bA_{k1}\bY_k)\ge(1-\epsilon_k^2/2)^2\\
\tr(\bA_{k2}\bY_k)=1\\
\tr(\bA_{k3}\bY_k)\le\beta_k^2\\
\bY_k\succeq\bzero.
 \end{array}
\end{array}
\end{equation}
and its dual is
\begin{equation}\label{the-worst-SINR-rewrite-qcqp-homog-matrix-dual}
\begin{array}[c]{ll}
\underset{ }{\sf{max}}  & \begin{array}[c]{c}(1-\epsilon_k^2/2)^2 x_{k1}+ x_{k2}+\beta_k^2 x_{k3} \end{array}\\
\sf{s.t.}               &
\begin{array}[t]{l}
\bA_{k0}-x_{k1}\bA_{k1}-x_{k2}\bA_{k2}-x_{k3}\bA_{k3}\succeq\bzero,\\
x_{k1}\ge0,\,x_{k2}\in\mathbb{R},\,x_{k3}\le0.
\end{array}
\end{array}
\end{equation}

We claim that the strong duality between \eqref{the-worst-SINR-rewrite-qcqp-homog-matrix-SDR} and \eqref{the-worst-SINR-rewrite-qcqp-homog-matrix-dual} holds. Given the strong duality result, we have
\begin{equation}\label{three-optimal-vals-equal}
v^\star(\eqref{the-worst-SINR-rewrite-qcqp-homog-matrix})=v^\star(\eqref{the-worst-SINR-rewrite-qcqp-homog-matrix-SDR})=v^\star(\eqref{the-worst-SINR-rewrite-qcqp-homog-matrix-dual}),
\end{equation} for $\epsilon_k\le \sqrt{2}$,  since there are three homogeneous constraint in \eqref{the-worst-SINR-rewrite-qcqp-homog-matrix-SDR} and a rank-one solution for it can always be constructed (see e.g. \cite{bookchapter-2010}).

\begin{proposition}\label{strong-duality-primal-dual-sdps}
It holds that both the two SDPs \eqref{the-worst-SINR-rewrite-qcqp-homog-matrix-SDR} and \eqref{the-worst-SINR-rewrite-qcqp-homog-matrix-dual} are strictly feasible and solvable\footnote{By saying ``solvable", it means that the problem is feasible and bounded below (for a minimization problem) and the
optimal value is attained (cf. \cite{Nemi-book2001}).}, and the optimal values of them are equal to each other.
\end{proposition}

See Appendix \ref{proof-strong-duality-primal-dual-sdps} for a proof.


Observing that $v^\star(\eqref{the-worst-SINR-rewrite-qcqp-2})=v^\star(\eqref{the-worst-SINR-rewrite-qcqp-homog-matrix})$, it follows from \eqref{three-problems-equal-values} and \eqref{three-optimal-vals-equal} that $v^\star(\eqref{the-worst-SINR-rewrite-qcqp})=v^\star(\eqref{the-worst-SINR-rewrite-qcqp-homog-matrix-dual})$.
Therefore, we can reexpress robust transmit beamforming problem \eqref{R-OBP-O-h-new-1-reformulation-1} as the following QMI problem:
\begin{subequations}\label{R-OBP-O-h-new-1-reformulation-1-QMI-full}
\begin{align} \label{R-OBP-O-h-new-1-reformulation-1-QMI-full.a}
\underset{}{\sf{min}}\,  & \sum_{k=1}^K\bw_k^H\bw_k \\ \label{R-OBP-O-h-new-1-reformulation-1-QMI-full.b}
\sf{s.t.}\,                               & (1-\epsilon_k^2/2)^2 x_{k1}+ x_{k2}+\beta_k^2 x_{k3}\ge\sigma_k^2\\ \label{R-OBP-O-h-new-1-reformulation-1-QMI-full.c}
 & \left[\begin{array}{cc}\tilde\bW_k-x_{k3}\bI&\sqrt{\alpha_k}x_{k3}\bI\\ \sqrt{\alpha_k}x_{k3}\bI&-x_{k1}\bh_{q_k}\bh_{q_k}^H-(x_{k2}+\alpha_k x_{k3})\bI\end{array}\right]\succeq\bzero\\ \label{R-OBP-O-h-new-1-reformulation-1-QMI-full.d}
 &\bw_k\in\mathbb{C}^{N_t},\,x_{k1}\ge0,\,x_{k2}\in\mathbb{R},\,x_{k3}\le0,\,\forall k,
\end{align}
\end{subequations}
where the matrix inequalities are equivalent to QMIs: $\bA_{k0}-x_{k1}\bA_{k1}-x_{k2}\bA_{k2}-x_{k3}\bA_{k3}\succeq\bzero$, $\forall k$.\footnote{Note that in the matrix inequality constraints of \eqref{R-OBP-O-h-new-1-reformulation-1-QMI-full}, there are quadratic terms $\bw_k\bw_k^H$, and that is why the matrix inequality is called a QMI (namely, all terms in the matrix inequality are quadratic or linear or a constant with respect to the optimization variables).}


 Recall that the equivalent QMI problem in \cite{MHMD2016tsp} (considering perturbation set \eqref{uncertainty-set-channel-2016-tsp} or \eqref{uncertainty-set-channel} with $\beta_k=0$) is
\begin{equation}\label{R-OBP-O-h-tsp2016}
\begin{array}[c]{ll}
\underset{}{\sf{min}}  & \begin{array}[c]{c}\sum_{k=1}^K\bw_k^H\bw_k \end{array}\\
\sf{s.t.}                               & \left[\begin{array}{cc}\tilde\bW_k+(x_{k1}+x_{k2})\bI&\tilde\bW_k\bh_{q_k}+x_{k2}\bh_{q_k}\\ \bh_{q_k}^H\tilde\bW_k+x_{k2}\bh_{q_k}^H&t_k\end{array}\right]\succeq\bzero\\
 & t_k=\bh_{q_k}^H\tilde\bW_k\bh_{q_k}-\sigma_k^2/\alpha_k-x_{k1}\epsilon_k^2\\
 &\bw_k\in\mathbb{C}^{N_t},\,x_{k1}\ge0,\,x_{k2}\in\mathbb{R},\forall k,
\end{array}
\end{equation}the SDP relaxation of which is problem (19) in \cite{MHMD2016tsp}.

\section{Restricted LMI Relaxation Problems}

Clearly, the conventional LMI relaxation technique can be applied to solve QMI problem \eqref{R-OBP-O-h-new-1-reformulation-1-QMI-full}:
\begin{subequations}\label{R-OBP-O-h-new-1-reformulation-1-QMI-full-LMI}
\begin{align}\label{R-OBP-O-h-new-1-reformulation-1-QMI-full-LMI.a}
\underset{}{\sf{min}}\,  & \sum_{k=1}^K\tr \bW_k \\ \label{R-OBP-O-h-new-1-reformulation-1-QMI-full-LMI.b}
\sf{s.t.}\,
 & \left[\begin{array}{cc}\hat\bW_k-x_{k3}\bI&\sqrt{\alpha_k}x_{k3}\bI\\ \sqrt{\alpha_k}x_{k3}\bI&-x_{k1}\bh_{q_k}\bh_{q_k}^H-(x_{k2}+\alpha_k x_{k3})\bI\end{array}\right]\succeq\bzero\\
 & \eqref{R-OBP-O-h-new-1-reformulation-1-QMI-full.b}\mbox{ satisfied}\\
 &\bW_k\succeq\bzero,\,x_{k1}\ge0,\,x_{k2}\in\mathbb{R},\,x_{k3}\le0,\,\forall k,
\end{align}
\end{subequations}
where $\hat\bW_k=\bW_k/\gamma_k-\sum_{j\ne k}\bW_j$.
 If the LMI problem has a rank-one solution $\{\bw_k^\star\bw_k^{\star H}\}$, then solution $\{\bw_k^\star\}$ is optimal for robust beamforming problem \eqref{R-OBP-O-h-new-1-reformulation-1}. If the LMI problem has a high-rank solution, we seek a restricted LMI relaxation problem for \eqref{R-OBP-O-h-new-1-reformulation-1-QMI-full} to solve it. 

 Observe from the second group constraints in \eqref{R-OBP-O-h-new-1-reformulation-1-QMI-full} that $\tilde\bW_k-x_{k3}\bI\succeq\bzero$, namely,
 \begin{equation}\label{suffi-1}
 \frac{1}{\gamma_k}\bw_k\bw_k^H-x_{k3}\bI\succeq\sum_{j\ne k}\bw_j\bw_j^H,\,\forall k.
 \end{equation}
Suppose that $\lambda$ and $\mu$ are the largest eigenvalues of the left-hand side and the right-hand side of \eqref{suffi-1}, respectively. Hence, it follows that $\lambda\ge\mu$, which is tantamount to  that
\begin{equation}\label{suffi-1-0}
\frac{1}{\gamma_k}\|\bw_k\|^2-x_{k3}\ge t_k,\,t_k\bI\succeq \sum_{j\ne k} \bw_j\bw_j^H,
\end{equation}
for some $t_k$. Plugging  \eqref{suffi-1-0} into \eqref{R-OBP-O-h-new-1-reformulation-1-QMI-full} gives
\begin{equation}\label{R-OBP-O-h-new-1-reformulation-1-QMI-full-plug-suffi-1-0}
\begin{array}[c]{ll}
\underset{}{\sf{minimize}}  & \begin{array}[c]{c}\sum_{k=1}^K\bw_k^H\bw_k \end{array}\\
\sf{subject\;to}               &
\begin{array}[t]{l}
\eqref{R-OBP-O-h-new-1-reformulation-1-QMI-full.b},\eqref{R-OBP-O-h-new-1-reformulation-1-QMI-full.c},\eqref{suffi-1-0} \mbox{ satisfied}\\
x_{k1}\ge0,x_{k3}\le0,\bw_k,x_{k2},t_k,\forall k.
\end{array}
\end{array}
\end{equation}
Evidently, \eqref{R-OBP-O-h-new-1-reformulation-1-QMI-full} is equivalent to \eqref{R-OBP-O-h-new-1-reformulation-1-QMI-full-plug-suffi-1-0} in terms that they share the optimal value and the set of optimal solutions.

The SDP relaxation of \eqref{suffi-1-0} includes
\begin{equation}\label{suffi-1-0-SDR-1-0}
\frac{1}{\gamma_k}\tr\bW_k-x_{k3}\ge t_k,
\end{equation}
and
\begin{equation}\label{suffi-1-0-SDR-1-1}
t_k\bI\succeq \sum_{j\ne k} \bW_j.
\end{equation}
Alternatively, the relaxation  of \eqref{suffi-1-0} contains \eqref{suffi-1-0-SDR-1-0} and
\begin{equation}\label{suffi-1-0-SDR-2-1}
\left[\begin{array}{ccccccc}t_k\bI&\bw_1&\cdots&\bw_{k-1}&\bw_{k+1}&\cdots&\bw_{K}\\
\bw_1^H&1&\cdots &0 &0 & \cdots & 0\\
\vdots&\vdots & \ddots&\vdots &\vdots &\ddots &\vdots \\
\bw_{k-1}^H& 0&\cdots &1&0 &\cdots &0 \\
\bw_{k+1}^H&0 &\cdots &0 &1 &\cdots &0 \\
\vdots&\vdots &\ddots &\vdots &\vdots &\ddots &\vdots \\
\bw_K^H&0 &\cdots &0 &0 &\cdots &1
\end{array}\right]\succeq\bzero,
\end{equation}
Therefore, the SDP relaxation problem for \eqref{R-OBP-O-h-new-1-reformulation-1-QMI-full-plug-suffi-1-0} is
\begin{equation}\label{R-OBP-O-h-new-1-reformulation-1-QMI-full-plug-suffi-1-0-sdr-1}
\begin{array}[c]{ll}
\underset{}{\sf{min}}  & \begin{array}[c]{c}\sum_{k=1}^K\tr\bW_k \end{array}\\
\sf{s.t}               &
\begin{array}[t]{l}
\eqref{R-OBP-O-h-new-1-reformulation-1-QMI-full.b},\eqref{R-OBP-O-h-new-1-reformulation-1-QMI-full-LMI.b},\eqref{suffi-1-0-SDR-1-0},\eqref{suffi-1-0-SDR-1-1} \mbox{ satisfied}\\
\bW_k\succeq\bzero,x_{k1}\ge0,x_{k3}\le0,x_{k2},t_k,\forall k,
\end{array}
\end{array}
\end{equation}
or
\begin{equation}\label{R-OBP-O-h-new-1-reformulation-1-QMI-full-plug-suffi-1-0-sdr-2}
\begin{array}[c]{ll}
\underset{}{\sf{min}}  & \begin{array}[c]{c}\sum_{k=1}^K\tr\bW_k \end{array}\\
\sf{s.t}               &
\begin{array}[t]{l}
\eqref{R-OBP-O-h-new-1-reformulation-1-QMI-full.b},\eqref{R-OBP-O-h-new-1-reformulation-1-QMI-full-LMI.b},\eqref{suffi-1-0-SDR-1-0},\eqref{suffi-1-0-SDR-2-1} \mbox{ satisfied}\\
\left[\begin{array}{cc}\bW_k&\bw_k\\ \bw_k^H& 1\end{array}\right]\succeq\bzero,x_{k1}\ge0,x_{k3}\le0,x_{k2},t_k,\forall k.
\end{array}
\end{array}
\end{equation}
It is seen that if SDP problem \eqref{R-OBP-O-h-new-1-reformulation-1-QMI-full-plug-suffi-1-0-sdr-1} (or \eqref{R-OBP-O-h-new-1-reformulation-1-QMI-full-plug-suffi-1-0-sdr-2}) has a rank-one solution $\{\bw_k^\star\bw_k^{\star H}\}$ (or $\{\bw_k^\star\bw_k^{\star H},\bw_k^\star\}$), then $\{\bw_k^\star\}$ is optimal for  \eqref{R-OBP-O-h-new-1-reformulation-1-QMI-full-plug-suffi-1-0}, i.e. for \eqref{R-OBP-O-h-new-1-reformulation-1-QMI-full}. Remark that there are more valid convex constraints in \eqref{R-OBP-O-h-new-1-reformulation-1-QMI-full-plug-suffi-1-0-sdr-1} or \eqref{R-OBP-O-h-new-1-reformulation-1-QMI-full-plug-suffi-1-0-sdr-2} than \eqref{R-OBP-O-h-new-1-reformulation-1-QMI-full-LMI}, and the SDR relaxation \eqref{R-OBP-O-h-new-1-reformulation-1-QMI-full-plug-suffi-1-0-sdr-1} or \eqref{R-OBP-O-h-new-1-reformulation-1-QMI-full-plug-suffi-1-0-sdr-2} appears tighter and thus is called a restricted SDP/LMI relaxation problem.\footnote{As for the worst-case computational complexity (see \cite[pages 423-424]{Nemi-book2001}), it is of $O(K^{3.5}N_t^{6.5})$ for problem \eqref{R-OBP-O-h-new-1-reformulation-1-QMI-full-plug-suffi-1-0-sdr-1}, and $O((K+2N_t)^{0.5}(5K^{3.5}N_t^{6}+K^{5.5}N_t^{4}+4K^{4.5}N_t^{5}))$ for problem \eqref{R-OBP-O-h-new-1-reformulation-1-QMI-full-plug-suffi-1-0-sdr-2}.}

\section{Numerical Examples}

We consider an MISO FDD downlink  system, where the BS is equipped with  eight antennas ($N_t=8$) and $K$ single-antenna users are served simultaneously. The channel from each BS antenna to each user is modeled as being independent with a circular complex Gaussian distribution with zero mean and unit variance. Each receiver's noise variance is fixed to $\sigma_k^2=0.01$. All the SINR thresholds are equal to 5~dB. The Grassmannian codebook downloaded from \url{https://engineering.purdue.edu/djlove/packings/two_d/} is employed. The norm of error $\bu_k$ (see \eqref{uncertainty-set-channel}) is bounded by $\beta_k$ taking value from $\{0,0.1,0.2,0.3,0.4\}$, and the  norm bound of error $\be_k$ is set to $\epsilon_k=0.04\times\sqrt{2}$ (since $\epsilon_k\le\sqrt{2}$), for each user $k$. A total of 2000 channel realizations are tested. 

Fig.~1 examines how the average transmit power and the problem feasibility rate are affected by the radius $\beta_k$, for both cases of five and six users ($K=5$ and $6$). When $\beta_k=0$, the SDP relaxation problem of \eqref{R-OBP-O-h-tsp2016} is solved; for $\beta_k\in\{0.1,0.2,0.3,0.4\}$, SDP problem \eqref{R-OBP-O-h-new-1-reformulation-1-QMI-full-LMI} is solved. We report that all the SDP problems have a rank-one solution as long as they are feasible. 
A practically logic result we observe from Fig.~1 is that more transmit power is required and the feasibility rate decreases as $\beta_k$ increases, and as the number of users is changed from five to six for a fixed $\beta_k$.

\begin{figure}
\centering
	\includegraphics{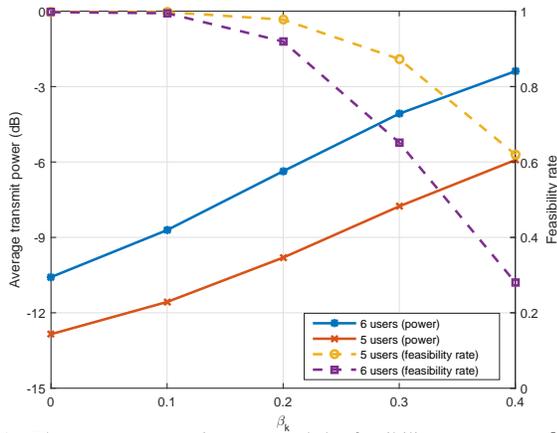}
\vspace{4.3cm}
	\caption{The average transmit power  and the feasibility rate versus $\beta_k$.}
	\label{fig-1}
\end{figure}

\begin{figure}
\centering
	\includegraphics{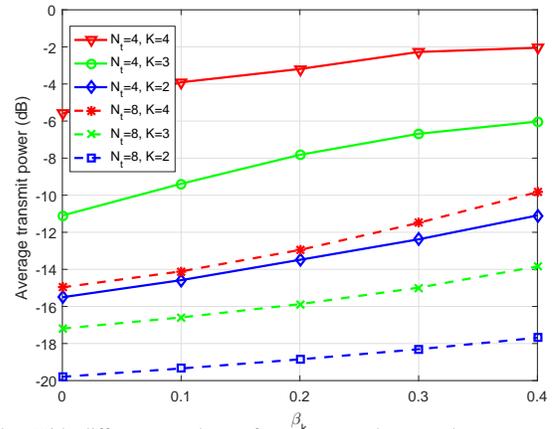}
\vspace{4.3cm}
	\caption{With different numbers of antennas and users, the average transmit power versus $\beta_k$.}
	\label{fig-2}
\end{figure}

With different numbers of antennas and users, Fig. \ref{fig-2} tests how the transmit power is impacted. As can be observed, when the number of users increases for the either four-antenna or eight-antenna case, more transmit power is required; when the number of antennas increases for a fixed $K$ (the number of users, and $K\in\{2,3,4\}$), less transmit power is sufficient. These observations are reasonable.

In order to check that either the restricted SDP problem \eqref{R-OBP-O-h-new-1-reformulation-1-QMI-full-plug-suffi-1-0-sdr-1} or \eqref{R-OBP-O-h-new-1-reformulation-1-QMI-full-plug-suffi-1-0-sdr-2} has a rank-one solution for the problem instances where the conventional SDP problem \eqref{R-OBP-O-h-new-1-reformulation-1-QMI-full-LMI} admits a solution of rank more than one, we create Table I, where the number of rank-one instances for \eqref{R-OBP-O-h-new-1-reformulation-1-QMI-full-plug-suffi-1-0-sdr-1}/number of rank-one instances for \eqref{R-OBP-O-h-new-1-reformulation-1-QMI-full-LMI}/number of feasible instances for \eqref{R-OBP-O-h-new-1-reformulation-1-QMI-full-LMI} are listed for the scenario of four transmit antennas ($N_t=4$) and three users ($K=3$). It is observed from the highlighted elements of the table that the restricted LMI relaxation problem \eqref{R-OBP-O-h-new-1-reformulation-1-QMI-full-plug-suffi-1-0-sdr-1} has a rank-one solution for the instances where  \eqref{R-OBP-O-h-new-1-reformulation-1-QMI-full-LMI} possesses a high-rank solution, except the case of $\beta_k=0.12$, $\epsilon_k=0.04\sqrt{2}$ and $\gamma_k=13$ dB. For this exceptional case, there is one problem instance where \eqref{R-OBP-O-h-new-1-reformulation-1-QMI-full-plug-suffi-1-0-sdr-1} has still a high-rank solution. Nevertheless, we report that the other restricted SDP relaxation problem \eqref{R-OBP-O-h-new-1-reformulation-1-QMI-full-plug-suffi-1-0-sdr-2} for the problem instance, has a rank-one solution. This implies that  SDP problem \eqref{R-OBP-O-h-new-1-reformulation-1-QMI-full-plug-suffi-1-0-sdr-1} or \eqref{R-OBP-O-h-new-1-reformulation-1-QMI-full-plug-suffi-1-0-sdr-2} can output a rank-one solution for the instances for which traditional  SDP problem \eqref{R-OBP-O-h-new-1-reformulation-1-QMI-full-LMI} has a high-rank solution.

\begin{table}[h]
\centering \caption{Occurrence of rank-one solutions for \eqref{R-OBP-O-h-new-1-reformulation-1-QMI-full-LMI} and \eqref{R-OBP-O-h-new-1-reformulation-1-QMI-full-plug-suffi-1-0-sdr-1}}
\label{tab-1}
\begin{tabular}{|c||c|c|}
  \hline
  $\beta_k$ & $\epsilon_k=0.04\sqrt{2}$, $\gamma_k=13$ dB & $\epsilon_k=0.08\sqrt{2}$, $\gamma_k=14$ dB \\
  \hline
  0.02 & {\bf 1834/1829/1834}  & {\bf 643/641/643} \\
  0.04 & {\bf 1747/1743/1747} & 415/415/415 \\
  0.06 & 1596/1596/1596 & 226/226/226 \\
  0.08 & 1410/1410/1410 & 108/108/108 \\
  0.10 & {\bf 1199/1197/1199} & 39/39/39 \\
  0.12 & {\bf 963/962/964} & 12/12/12 \\
  0.14 & 736/736/736 & 3/3/3 \\
  0.16 & 544/544/544 & 0/0/0 \\
  0.18 & 362/362/362 & 0/0/0 \\
  0.20 & 238/238/238 & 0/0/0 \\
  \hline
\end{tabular}
\end{table}


\section{Conclusion}

In an FDD  downlink system, we have considered a robust MISO beamforming optimization problem of the transmit power minimization subject to robust SINR constraints. Suppose that the perturbation of the actual channel vector is caused by the channel direction quantization errors, channel estimation errors and outdated channel effects. Then, the robust SINR constraints have been transformed into QMI constraints with respect to the beamforming vectors by resorting to an equivalent phase rotation process and the strong duality of QCQP, and the robust beamforming problem has been reformulated into a QMI problem. Finally, the problem has been solved by a restricted LMI relaxation technique.

\appendix


%
%

\subsection{Proof of Proposition \ref{strong-duality-primal-dual-sdps}}\label{proof-strong-duality-primal-dual-sdps}

Let $(x_{k1},x_{k2},x_{k3})=(1,\alpha_k\gamma,\gamma)$ with $\gamma<0$. It is seen that $-x_2\bA_{k2}-x_{k3}\bA_{k3}\succ\bzero$ for any $\gamma<0$, and therefore, $\bA_{k0}-x_{k1}\bA_{k1}-x_2\bA_{k2}-x_{k3}\bA_{k3}\succ\bzero$ for sufficiently small $\gamma$. In other words, dual SDP \eqref{the-worst-SINR-rewrite-qcqp-homog-matrix-dual} is strictly feasible.

Note that $(\sqrt{\alpha_k}\bh_{q_k},\bh_{q_k})$ is strictly feasible for \eqref{the-worst-SINR-rewrite-qcqp-2}, since $\epsilon_k>0$. One defines the following matrix
\begin{equation}
\bY(\lambda)=\lambda\left[\begin{array}{c}\sqrt{\alpha_k}\bh_{q_k}\\ \bh_{q_k} \end{array}\right]\left[\begin{array}{c}\sqrt{\alpha_k}\bh_{q_k}\\ \bh_{q_k} \end{array}\right]^H+(1-\lambda)\left[\begin{array}{cc}\bI&\bzero\\ \bzero&\frac{1}{N_t}\bI\end{array}\right],
\end{equation}
for $\lambda\in(0,1)$. Clearly, $\bY(\lambda)\succ\bzero$ for any $\lambda\in(0,1)$.
It is checked that $\tr(\bA_{k2}\bY(\lambda))=1$ for $\lambda\in(0,1)$, and when $\lambda$ is sufficiently close to one, we have that $\tr(\bA_{k1}\bY(\lambda))>(1-\epsilon_k^2/2)^2$ and $\tr(\bA_{k3}\bY(\lambda))<\beta_k^2$. Therefore, $\bY(\lambda)$ is a strictly feasible solution for \eqref{the-worst-SINR-rewrite-qcqp-homog-matrix-SDR} when  $\lambda$ is sufficiently close to one.

Since both the primal SDP and the dual SDP are strictly feasible, it hence follows from \cite[Theorem 2.4.1]{Nemi-book2001} that both of them are solvable and the optimal values are equal to each other.





\end{document}